\documentclass[twocolumn,useAMS,usenatbib]{mnras}

\usepackage{graphicx}
\usepackage{subfigure}
\usepackage{xcolor}
\usepackage{mathtext,bbm,amsmath,amsfonts,amssymb,indentfirst,syntonly,graphicx}
\usepackage{mathtools}
\usepackage{slashbox}
\usepackage[english]{babel}
\usepackage{calc}
\usepackage{tikz}
\usepackage[T1]{fontenc}
\usepackage{ae,aecompl}

\def\bc{\begin{center}}
\def\ec{\end{center}}
\def\be{\begin{eqnarray}}
\def\ee{\end{eqnarray}}

\title[GRB polarization reduction induced by LIV]{Gamma-ray burst polarization reduction induced by the Lorentz invariance violation}
\author[H.-N. Lin, X. Li and Z. Chang]
{Hai-Nan Lin$^{1}$\thanks{e-mail: linhn@ihep.ac.cn.}, Xin Li$^{1,2}$\thanks{e-mail: lixin1981@cqu.edu.cn.}, Zhe Chang$^{3}$\\
$^{1}$Department of Physics, Chongqing University, Chongqing 401331, China\\
$^{2}$State Key Laboratory of Theoretical Physics, Institute of Theoretical Physics, Chinese Academy of Sciences, Beijing 100190, China\\
$^{3}$Institute of High Energy Physics, Chinese Academy of Sciences, Beijing 100049, China\\}
\begin{document}

\date{Accepted xxxx; Received xxxx; in original form xxxx}

\pagerange{\pageref{firstpage}--\pageref{lastpage}} \pubyear{2016}

\maketitle

\label{firstpage}

\begin{abstract}
  It has been observed that photons in the prompt emission of some gamma-ray bursts (GRBs) are highly polarized. The high polarization is used by some authors to give a strict constraint on the Lorentz invariance violation (LIV). If the Lorentz invariance is broken, the polarization vector of a photon may rotate during its propagation. The rotation angle of polarization vector depends on both the photon energy and the distance of source. It is believed that if high polarization is observed, then the relative rotation angle (denoted by $\alpha$) of polarization vector of the highest energy photon with respect to that of the lowest energy photon should be no more than $\pi/2$. Otherwise, the net polarization will be severely suppressed, thus couldn't be as high as what was actually observed. In this paper, we will give a detailed calculation on the evolution of GRB polarization arising from LIV effect duration the propagation. It is shown that the polarization degree rapidly decrease as $\alpha$ increases, and reaches a local minimum at $\alpha\approx \pi$, then increases until $\alpha\approx 3\pi/2$, after that decreases again until $\alpha \approx 2\pi$, etc. The polarization degree as a function of $\alpha$ oscillates with a quasi-period $T\approx \pi$, while the oscillating amplitude gradually decreases to zero. Moreover, we find that a considerable amount (more than $60\%$ of the initial polarization) of polarization degree can be conserved when $\alpha\approx \pi/2$. The polarization observation in a higher and wider energy band, a softer photon spectrum, and a higher redshift GRB is favorable in order to tightly constrain LIV effect.
\end{abstract}

\begin{keywords}
polarization -- gamma-ray burst: general
\end{keywords}

\section{Introduction}\label{sec:introduction}

Lorentz invariance is one of the foundations of Einstein's special relativity. It has been tested to a high accuracy using both the laboratory and cosmos experiments. In some quantum gravity theories \citep{Kostelecky:1989,Gambini:1999,Amelino-Camelia:2002,Myers:2003}, however, Lorentz invariance may be broken. In such a case, the propagation of light in vacuum exhibits a nontrivial dispersion relation compared to that in the special relativity. One of the most extensively discussed dispersion relation with Lorentz invariance violation (LIV) is $E_{\pm}^2=p^2\pm 2\xi p^3/M_{\rm pl}$, where $M_{\rm pl}$ is the Planck energy, and $\xi$ is a dimensionless parameter. According to this dispersion relation, the group velocity of light in vacuum, $v_g=\partial E/\partial p$, is no longer a constant, but is energy dependent. Therefore, two photons with different energies emitted simultaneously from a cosmological source will have a slightly time delay when they arrive the earth. Such a time delay should be detectable if the source is far enough away from the earth. Gamma-ray bursts (GRBs) provide an effect tool to test Lorentz invariance. As one of the most energetic explosions in the universe, GRBs are detectable out to redshift $z\approx 10$.  In fact, GRBs have been widely used to constrain LIV effect \citep{Ellis:2006,Jacob:2008,Abdo:2009,Chang:2012,Nemiroff:2012fk,Zhang:2014wpb,Vasileiou:2015}. The value of $\xi$ constrained in this way is usually in the order of unity. The strictest limit on the LIV energy scale from GRB 090510 is $E_{\rm QG}>7.43\times 10^{21}$ GeV \citep{Nemiroff:2012fk}, which corresponds to $\xi<1.6\times 10^{-3}$.

Much tighter constraints can be obtained through the measurement of GRB polarization. The polarimetric observations show that photons in the prompt emission of some GRBs are highly linearly polarized. For example, The polarization degree of the first reported highly polarized burst, GRB 021206, is about $80\%\pm 20\%$ \citep{Coburn:2003}. However, a following re-analysis of the same data found no significant polarization signal \citep{Rutledge:2004}. The highest polarized burst reported so far is GRB 041219A, which has polarization degree $98\% \pm 33\%$ \citep{Kalemci:2007}. But again this result was criticized by a more detailed analysis \citep{McGlynn:2007}. In 2011, the gamma-ray burst polarimeter {\it GAP} \citep{Yonetoku:2011a} onboard the Japanese Interplanetary Kite-craft Accelerated by Radiation Of the Sun (IKAROS) detected two highly polarized bursts, GRB 110301A and 110721A. These two bursts have conform polarization degree of $70\%\pm 22\%(3.7\sigma)$ and $84_{-28}^{+16}\%(3.3\sigma)$, respectively \citep{Yonetoku:2012}. The temporal evolution of polarization has also been observed \citep{Greiner:2003,Gotz:2009,Yonetoku:2011b,Mundell:2013}. Although the uncertainty is still large and many controversies exist, the possibility that some GRBs are highly polarized can't be excluded. The high accuracy $\gamma$-ray polarimeter {\it POLAR} \citep{Xiao:2015} onboard the Chinese space laboratory Tiangong-II is fully designed to measure the GRB polarization in $50-500$ keV energy band. It is scheduled to launch in September, 2016. If {\it POLAR} is launched, the GRB polarimetric data will be significantly enlarged, and the statistical significance will be highly improved. In the theoretical aspect, several theoretical models have been proposed to explain the GRB polarization \citep{Sari:1999,Waxman:2003,Granot:2003dy,Lazzati:2004,Toma:2009,Mao:2013gha,Chang:2014,Chang:2014a,Chang:2014b,Lan:2016}.

If GRBs are really highly polarized, it will give a strict constraint on LIV effect. The idea of using polarization to constrain LIV was first proposed by \citet{Gleiser:2001rm}. They analysed the polarimetric data in ultraviolet band for radio galaxy 3C 256 locating at a redshift of 1.82, and obtained $\xi<10^{-4}$. When Lorentz invariance is broken, the polarization vector (i.e., the electric component) of a photon will rotate during its propagation. Suppose a beam of photons emit from a GRB source and propagate to the observer on earth. Every photon will rotates its polarization vector by an angle $\Delta \theta(k)$, which depends on the photon energy. Let $\alpha$ to be the difference of rotation angles of polarization vectors between the highest energy photon and the lowest energy photon. If high polarization degree is observed, then $\alpha$ couldn't be too large, otherwise the net polarization will be severely suppressed. \citet{Toma:2012} set the upper limit of $\alpha$ to be $\pi/2$, and obtained a strict upper limit on the value $\xi$ in the order of $\mathcal{O}(10^{-15})$ from the polarimetric data of three GRBs. However, the GRBs used by \citet{Toma:2012} have no direct measurement of redshift, while redshift deduced from the empirical luminosity correlations has large uncertainty. \citet{Gotz:2013} used the polarization data of GRB 061122 locating at a redshift $z=1.33$, and obtained $\xi<3.4\times 10^{-16}$. Using the most distant polarized burst, GRB 140206A, which has a confirm redshift measurement of $z=2.739$, \citet{Gotz:2014vza} have obtained the strictest constraint to date, i,e., $\xi<1\times 10^{-16}$. All of these constraints are based on the assumption that the rotation angle $\alpha$ is smaller than $\pi/2$. Then a question arises: how much polarization degree can be conserved if the polarization vector changes an angle $\alpha$\,? The main aim of our paper is to address this question. We will give a detailed calculation on the evolution of GRB polarization as a function of $\alpha$, and show that a considerable amount of polarization (depending on the photon energy band) can be conserved even if $\alpha$ is larger than $\pi/2$.

The rest parts of this paper are arranged as follows: In Section \ref{sec:general}, we present the general formulae for the evolution of polarization induced by LIV effect. In Section \ref{sec:special}, we employ the formulae to GRBs, which usually have power-law spectra in the energy band of $\sim$ keV. Three different cases are discussed: (1) photons are initially completely unpolarized, (2) photons are initially completely polarized, and (3) photons are initially partially polarized. Finally, discussions and conclusions are given in Section \ref{sec:conclusions}.

\section{General formulae}\label{sec:general}

One of the most discussed, Lorentz invariance violating dispersion relation of photon can be parameterized as \citep{Myers:2003}
\begin{equation}\label{eq:dispersion}
  E_{\pm}^2=p^2\pm 2\xi p^3/M_{\rm pl},
\end{equation}
where ``\,$\pm$\," represents the left-handed or right-handed states, $M_{\rm pl}\approx 1.22\times 10^{19}$ GeV is the Planck energy, and $\xi$ is a dimensionless parameter characterizing the magnitude of LIV effect. The helicity dependence of photon velocity will lead to the vacuum birefringence effect, such that the polarization vector of a photon will rotate during the propagation. The rotation angle of polarization vector for a photon with energy $p$ (in unit of $\hbar=c=1$) propagating from the source at redshift $z$ to the observer on earth is given by \citep{Laurent:2011he,Toma:2012}
\begin{equation}\label{eq:delta-theta}
  \Delta\theta(k)=\xi\frac{k^2F(z)}{M_{\rm pl}H_0},
\end{equation}
where $k=p/(1+z)$ is the observed photon energy, $H_0$ is the Hubble constant, and
\begin{equation}
  F(z)=\int_0^z\frac{(1+z)dz}{\sqrt{\Omega_M(1+z)^3+\Omega_{\Lambda}}}.
\end{equation}
Throughout this paper, we take $H_0=68~{\rm km}\,{\rm s}^{-1}\,{\rm Mpc}^{-1}$, $\Omega_M=0.3$ and $\Omega_{\Lambda}=0.7$ from the Planck 2015 results \citep{Ade:2015xua}.

Suppose a beam of non-coherent light emits from the source. Choose the propagation direction as the $z$-axis, and the polarization direction is in the $xy$-plane. The intensity of photons whose electric vector is in the infinitesimal azimuth angle interval $d\theta$, and whose energy is in the infinitesimal interval $dk$ can be written as
\begin{equation}\label{eq:intensity1}
  dj(\theta,k)=j_0f(\theta)kN(k)d\theta dk,
\end{equation}
where $N(k)$ is the photon number spectrum, $f(\theta)$ is a periodic function of $\theta$ with period $\pi$, and $j_0$ is a normalization constant. Since photon intensity is proportional to the square of electric vector, the intensity projected onto the direction of azimuth angle $\varphi$ is given by
\begin{equation}\label{eq:intensity2}
  dj_{\varphi}(\theta,k)=j_0f(\theta)kN(k)\cos^2(\varphi-\theta)d\theta dk.
\end{equation}
Therefore, the total intensity of photons polarized along the direction $\varphi$ is given by
\begin{equation}\label{eq:intensity3}
  I(\varphi)=\int dj_{\varphi}(\theta,k)=\int_0^{\pi}d\theta\int_{k_1}^{k_2}dk~j_0f(\theta)kN(k)\cos^2(\varphi-\theta),
\end{equation}
where $k_1$ and $k_2$ are the lower and upper limits of photon spectrum, respectively. The polarization degree is defined by \citep{Rybicki:1979}
\begin{equation}\label{eq:polarization1}
  \Pi=\frac{I_{\rm max}-I_{\rm min}}{I_{\rm max}+I_{\rm min}},
\end{equation}
where $I_{\rm max}$ and $I_{\rm min}$ are the maximum and minimum of $I(\varphi)$, respectively.

As photons propagate from the source to the observer on earth, every photon will change its polarization vector according to equation (\ref{eq:delta-theta}).
Let $\alpha\equiv\Delta\theta(k_2)-\Delta\theta(k_1)$ to be the rotation angle of the electric vector of the highest energy photon with respect to that of the lowest energy photon, then equation (\ref{eq:delta-theta}) can be rewritten as
\begin{equation}\label{eq:delta-theta2}
  \Delta\theta(k)=\alpha\frac{k^2}{k_2^2-k_1^2}.
\end{equation}
The received photon intensity can be obtained from equation (\ref{eq:intensity3}) by replacing $f(\theta)$ with $f(\theta+\Delta\theta(k))$, i.e.,
\begin{equation}\label{eq:intensity4}
  I'(\varphi)=\int_0^{\pi}d\theta\int_{k_1}^{k_2}dk~j_0f(\theta+\Delta\theta(k))kN(k)\cos^2(\varphi-\theta).
\end{equation}
Therefore, the observed polarization degree is given by
\begin{equation}\label{eq:polarization2}
  \Pi'=\frac{I'_{\rm max}-I'_{\rm min}}{I'_{\rm max}+I'_{\rm min}},
\end{equation}
where $I'_{\rm max}$ and $I'_{\rm min}$ are the maximum and minimum of $I'(\varphi)$, respectively.

As a trivial example, let us consider the monochromatic photons. In this case, the photon spectrum can be written as the Dirac $\delta$-function
\begin{equation}\label{eq:mono-spectrum}
  N(k)=\delta(k-k_0).
\end{equation}
Substituting equation (\ref{eq:mono-spectrum}) into equation (\ref{eq:intensity3}), we obtain the initial light beam
\begin{eqnarray}\label{eq:mono-intensity1}
  I(\varphi)&=&\nonumber \int_0^{\pi}d\theta\int_{k_1}^{k_2}dk~j_0f(\theta)k\delta(k-k_0)\cos^2(\varphi-\theta)\\
  &=&j_0 k_0 \int_0^{\pi}d\theta f(\theta)\cos^2(\varphi-\theta).
\end{eqnarray}
Substituting equation (\ref{eq:mono-spectrum}) into equation (\ref{eq:intensity4}), we obtain the received light beam
\begin{eqnarray}\label{eq:mono-intensity2}
  I'(\varphi)&=&\nonumber \int_0^{\pi}d\theta\int_{k_1}^{k_2}dk~j_0k\delta(k-k_0)\\
  & &\nonumber \times f(\theta+\Delta\theta(k))\cos^2(\varphi-\theta)\\
  &=& j_0 k_0 \int_0^{\pi}d\theta f(\theta+\Delta\theta(k_0))\cos^2(\varphi-\theta).
\end{eqnarray}
Noticing $f(\theta+\pi)=f(\theta)$, through a simple calculation we can easily verify $I'(\varphi)=I(\varphi+\Delta\theta(k_0))$. This means that $I'$ is identical to $I$ except that the polarization direction is shifted by an angle $\Delta\theta(k_0)$, while the polarization degree does not change. This is in our expectation, because all the photons change the same polarization angle when the light beam is monochromatic.

\section{Power-law photons}\label{sec:special}

In this section, we focus on discussing the polarization evolution of GRB photons. The photon spectrum of a typical GRB in a wide energy range (from a few keV to tens MeV) can be well described by the Band function \citep{Band:1993}, which is a two smoothly jointed power laws cutting at a breaking energy. In the keV energy band, in which the polarization is measured, the Band function is well approximated by two power laws, i.e., $N(k)\propto k^{-\alpha}$ for $k<k_0$ and $N(k)\propto k^{-\beta}$ for $k>k_0$, where $k_0$ is the breaking energy, $\alpha$ and $\beta$ are the power-law indices. For a typical GRB, $\alpha \approx 1$ and $\beta\approx 2.2$ \citep{Preece:2000fv}. The working energy band of most $\gamma$-ray polarimeters at present is very narrow. In such a narrow energy band, the spectrum can even be well modelled by the simple power law, i.e., $N(k)\propto k^{-p}$. For simplicity, in the following we assume that the GRB spectrum is the simple power law, and the index $p$ can vary from $1.0$ to $2.5$. Three initial polarization states are considered: (1) initially completely unpolarized, (2) initially completely polarized, and (3) initially partially polarized.

{\bf Initially completely unpolarized} If photon beam is initially completely unpolarized, then $f(\theta)$ is independent of $\theta$. Without loss of generality, we take $f(\theta)=1$. Using $N(k)=k^{-p}$ and $f(\theta)=1$, equation (\ref{eq:intensity3}) simplifies to
\begin{eqnarray}
  I(\varphi)&=&\nonumber \int_0^{\pi}d\theta\int_{k_1}^{k_2}dk~j_0 k^{-p+1}\cos^2(\varphi-\theta)\\
  &=&\frac{\pi}{2}j_0\int_{k_1}^{k_2}dk~k^{-p+1}.
\end{eqnarray}
Similarly, equation (\ref{eq:intensity4}) simplifies to
\begin{eqnarray}
  I'(\varphi)&=&\nonumber \int_0^{\pi}d\theta\int_{k_1}^{k_2}dk~j_0 k^{-p+1}\cos^2(\varphi-\theta)\\
  &=&\frac{\pi}{2}j_0\int_{k_1}^{k_2}dk~k^{-p+1}.
\end{eqnarray}
We may see that $I'$ is identical to $I$, and both are independent of $\varphi$. Hence, initially unpolarized photons are always unpolarized duration the propagation.

{\bf Initially completely polarized} If photon beam is initially completely polarized, e.g., along the $x$-axis, then $f(\theta)$ can be written as\footnote{The summation of $\delta$-functions ensures that $f(\theta)$ is periodic, i.e., $f(\theta+\pi)=f(\theta)$.}
\begin{equation}\label{eq:f-theta}
  f(\theta)=\sum_{n=-\infty}^{+\infty}\delta(\theta-n\pi).
\end{equation}
Substituting $N(k)=k^{-p}$ and equation (\ref{eq:f-theta}) into equation (\ref{eq:intensity3}), we obtain
\begin{eqnarray}
  I(\varphi)&=& \nonumber \int_0^{\pi}d\theta\int_{k_1}^{k_2}dk~j_0 k^{-p+1}\\
  & &\nonumber \times \sum_{n=-\infty}^{+\infty}\delta(\theta-n\pi)\cos^2(\varphi-\theta)\\
  &=&j_0\cos^2\varphi \int_{k_1}^{k_2}dk~k^{-p+1}.
\end{eqnarray}
We can easily see that $I_{\rm max}=I(0)$ and $I_{\rm min}=I(\pi/2)=0$, which means that the initial polarization degree is $100\%$. Substituting $N(k)=k^{-p}$ and equations (\ref{eq:delta-theta2}) and (\ref{eq:f-theta}) into equation (\ref{eq:intensity4}), we obtain the observed photon intensity
\begin{eqnarray}
  I'(\varphi)&=&\nonumber \int_0^{\pi}d\theta\int_{k_1}^{k_2}dk~j_0k^{-p+1}\\
  & &\nonumber \times \sum_{n=-\infty}^{+\infty}\delta(\theta+\frac{\alpha k^2}{k_2^2-k_1^1}-n\pi)\cos^2(\varphi-\theta)\\
  &=&j_0 \int_{k_1}^{k_2}dk~k^{-p+1}\cos^2(\varphi+\frac{\alpha k^2}{k_2^2-k_1^1}).
\end{eqnarray}

\begin{figure}
  \centering
  \includegraphics[width=0.5\textwidth]{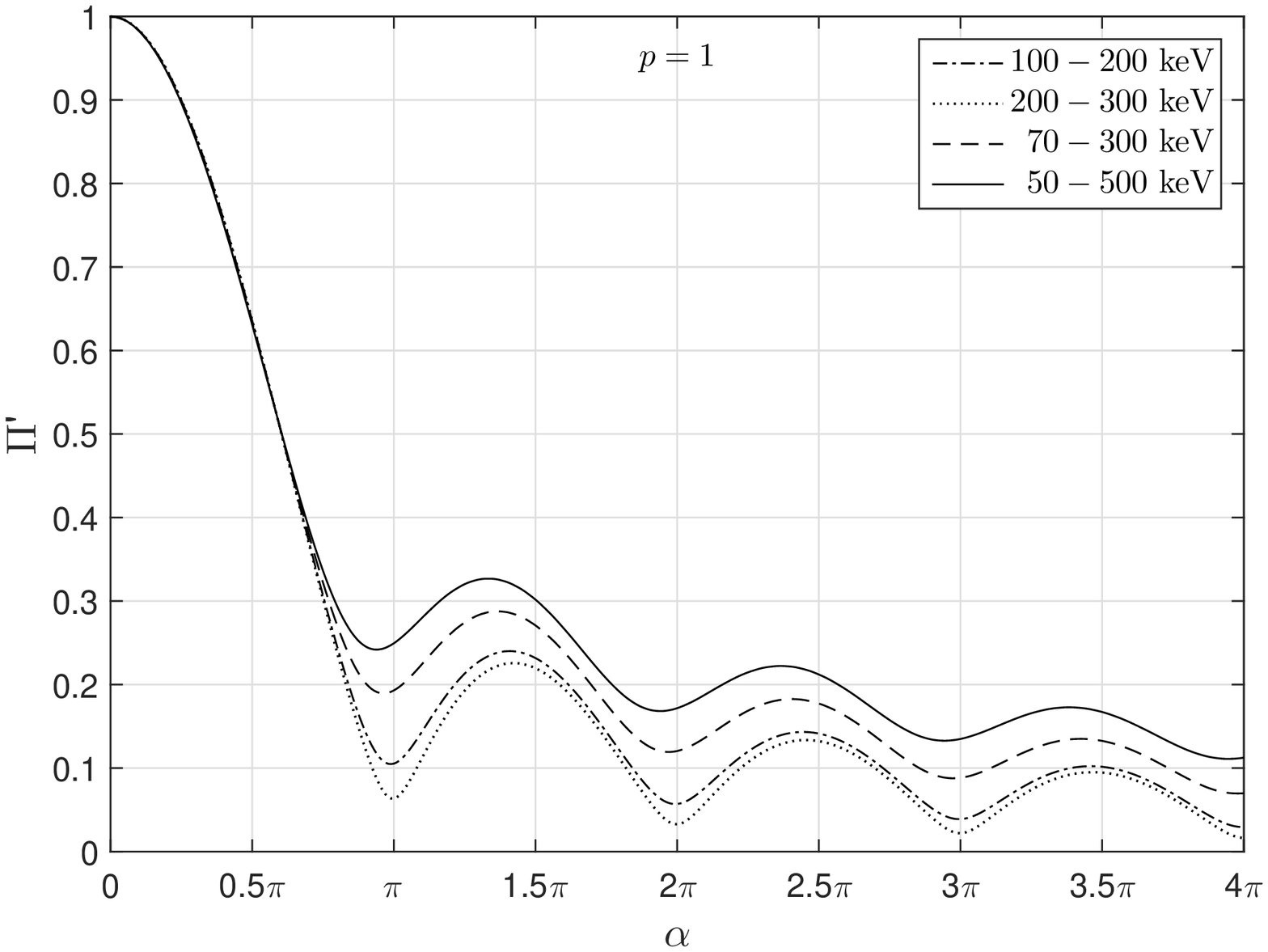}
  \includegraphics[width=0.5\textwidth]{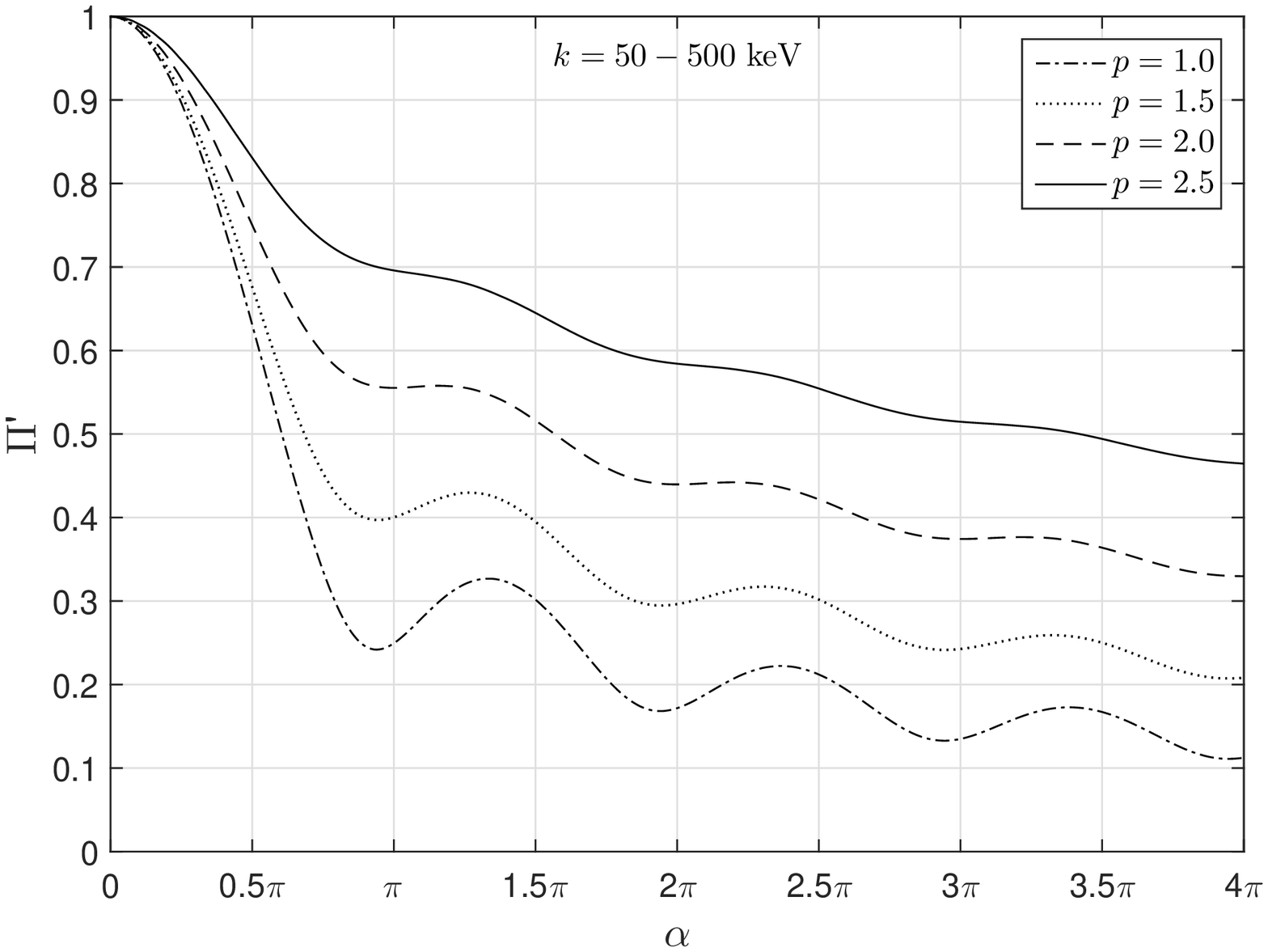}
  \caption{\small{Polarization degree as a function of $\alpha$ in the initially completely polarized case. Upper panel: the power-law index is fixed at $p=1$. Lower panel: the photons are in the energy band $k\in [50,500]$ keV.}}\label{fig:PI_alpha1}
\end{figure}

For any given $\alpha$, we can numerically calculate the maximum and minimum values of $I'(\varphi)$, then calculate the polarization degree according to equation (\ref{eq:polarization2}). We plot the polarization degree as a function of $\alpha$ in Figure \ref{fig:PI_alpha1}.
In the upper panel of Figure \ref{fig:PI_alpha1}, the power-law index is fixed at $p=1$. Curves for four energy bands are shown: $100-200$ keV, $200-300$ keV, $70-300$ keV and $50-500$ keV. The last two are the working energy bands of the {\it IKAROS-GAP}, and the {\it POLAR}, respectively. The polarization degree $\Pi'$ decreases rapidly as $\alpha$ increases, and reaches a local minimum at $\alpha\approx \pi$. After that, $\Pi'$ increases until $\alpha\approx 3\pi/2$, then decreases again until $\alpha \approx 2\pi$, etc. The polarization degree as a function of $\alpha$ oscillates with a quasi-period $T\approx \pi$, and the oscillating amplitude gradually decreases to zero. In the range $\alpha<\pi/2$, polarization degree is almost independent of photon energy. More than $60\%$ of the initial polarization is conserved at $\alpha=\pi/2$. When $\alpha>\pi$, the energy dependence of polarization is evident. Comparing the dot-dashed ($100-200$ keV) and dotted ($200-300$ keV) lines, we may see that for the same energy width, high energy photons have smaller polarization than low energy photons at the fixed rotation angle $\alpha$. This is because high energy photons rotate a larger absolute polarization angle $\Delta\theta$ (see equation (\ref{eq:delta-theta})) than low energy photons propagating the same distance. Therefore, the polarization direction of high energy photons are more mixed and the polarization degree are suppressed. Comparing the dotted ($200-300$ keV), dashed ($70-300$ keV) and solid ($50-500$ keV) lines, we may see that at a fixed rotation angle, photons in a wider energy band have larger polarization.

To see the possible dependence of polarization degree on the power-law index, we plot in the lower panel of Figure \ref{fig:PI_alpha1} for different values of $p$, while the photons are assumed to be in the energy band $k\in [50,500]$ keV. It is clearly shown that the net polarization degree increases as $p$ increase. This is easy to understand. When $p$ is larger, the photons are closer to monochromatic. While we have already showed that in the strictly monochromatic case, polarization degree does not change during the propagation. Therefore, to tightly constrain the LIV effect, GRBs with soft spectrum (i.e. smaller $p$ value) are favored.

\begin{figure}
 \centering
 \includegraphics[width=0.5\textwidth]{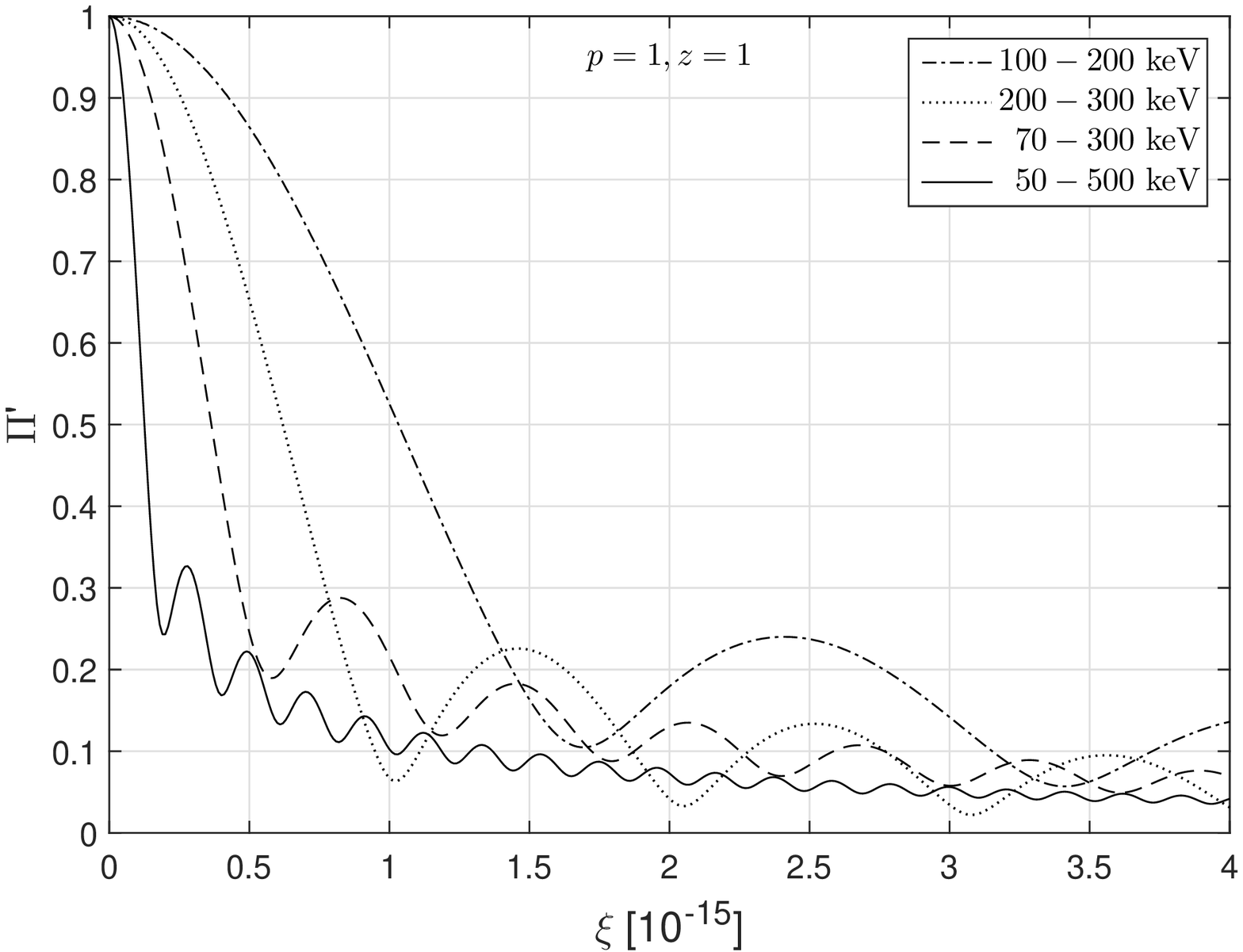}
 \includegraphics[width=0.5\textwidth]{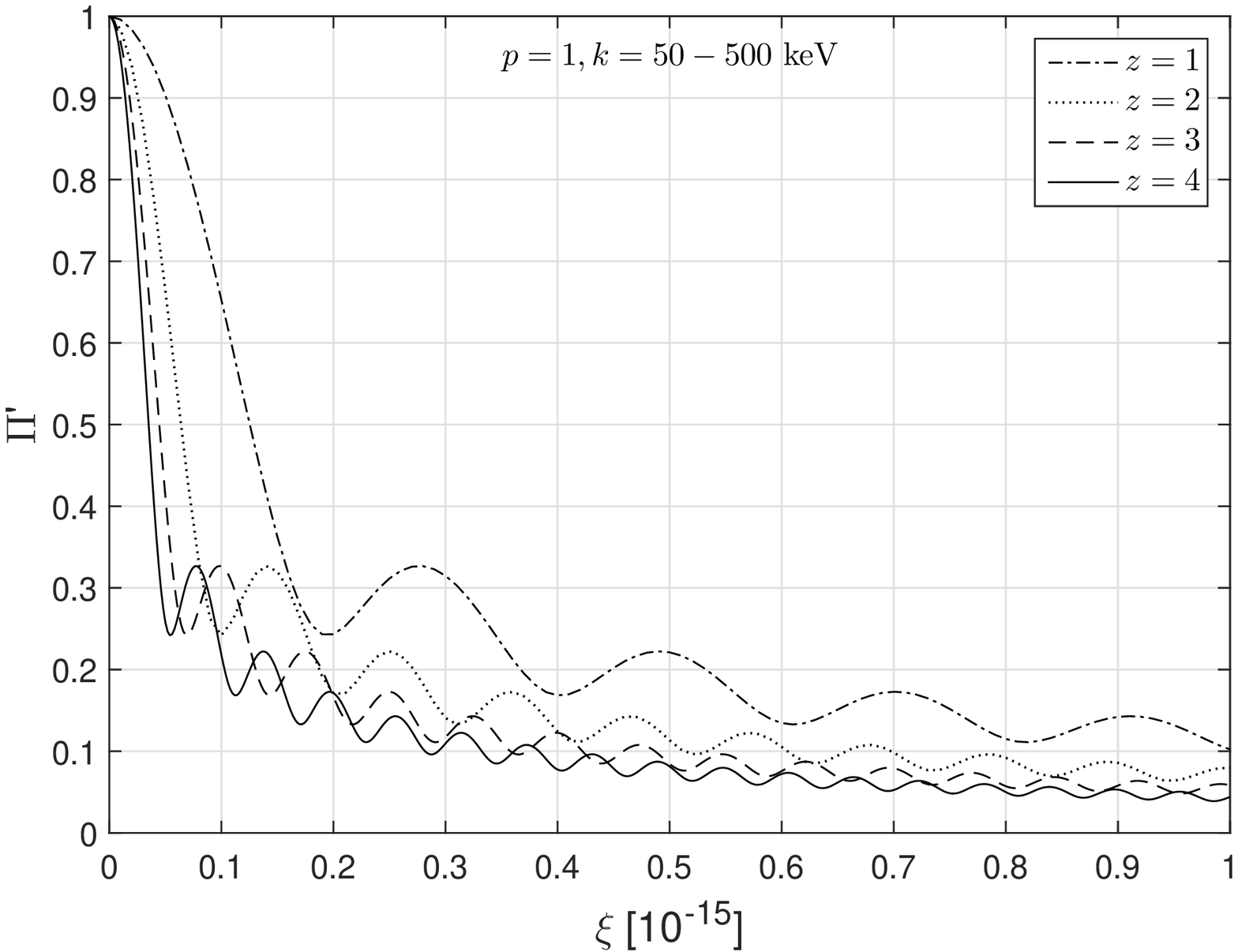}
 \caption{\small{Polarization degree as a function of $\xi$ in the initially completely polarized case. Upper panel: the redshift of source is $z=1$. Lower panel: the photons are in the energy band $k\in [50,500]$ keV. In both panels, the power-law index is $p=1$.}}\label{fig:PI_xi1}
\end{figure}

Although at the fixed rotation angle, photons in a wider energy band have larger polarization degree, it does not mean that we can indeed see larger polarization in a wider energy band. This is because a wider energy band will lead to a larger rotation angle. To see this more clearly, we plot, in the upper panel of Figure \ref{fig:PI_xi1}, the polarization as a function of $\xi$ in four energy bands.
The power-law index is taken to be $p=1$, and the redshift of the source is assumed to be $z=1$. Similar to the $\Pi'-\alpha$ plot, the $\Pi'-\xi$ plot also shows oscillation. However, unlike the $\Pi'-\alpha$ plot, in which the oscillating period is almost independent of photon energy, in the $\Pi'-\xi$ plot a higher and wider energy band has a smaller period. At the beginning, the polarization degree decreases rapidly as $\xi$ increases. A higher and wider energy band shows a steeper slope. At large $\xi$, where the net polarization is below $20\%$, curves of different energy bands intersect with each other complexly. To see the redshift dependence, in the lower panel of Figure \ref{fig:PI_xi1}, we plot the polarization degree as a function of $\xi$ for various redshifts. As is expected, the polarization degree decreases more rapidly at higher redshift.

{\bf Initially partially polarized} If photon beam is initially partially polarized, the concrete form of $f(\theta)$ depends on the initial polarization degree. We consider that photons are initially half polarized. In this case, we may choose $f(\theta)=\cos^2\theta$. Using $N(k)=k^{-p}$ and $f(\theta)=\cos^2\theta$, equation (\ref{eq:intensity3}) simplifies to
\begin{eqnarray}
  I(\varphi)&=& \nonumber \int_0^{\pi}d\theta\int_{k_1}^{k_2}dk~j_0k^{-p+1}\cos^2\theta\cos^2(\varphi-\theta)\\
  &=&\frac{1}{8}\pi (2+\cos2\varphi) j_0 \int_{k_1}^{k_2}dk~k^{-p+1}.
\end{eqnarray}
We can see that $I_{\rm max}=I(0)$ and $I_{\rm min}=I(\pi/2)$. The initial polarization degree, according to the definition in equation (\ref{eq:polarization1}), is $\Pi=50\%$. Substituting $N(k)=k^{-p}$ and $f(\theta)=\cos^2\theta$ into equation (\ref{eq:intensity4}), and noticing that $\Delta\theta(k)$ is given by equation (\ref{eq:delta-theta2}), we obtain
\begin{eqnarray}
  I'(\varphi)&=& \nonumber \int_0^{\pi}d\theta\int_{k_1}^{k_2}dk~j_0k^{-p+1}\\
  & & \nonumber \times \cos^2(\theta+\frac{\alpha k^2}{k_2^2-k_1^2})\cos^2(\varphi-\theta)\\
  &=&\frac{1}{8}\pi j_0 \int_{k_1}^{k_2}dk~k^{-p+1}[2+\cos(2\varphi+\frac{2\alpha k^2}{k_2^2-k_1^2})].
\end{eqnarray}

\begin{figure}
 \centering
 \includegraphics[width=0.5\textwidth]{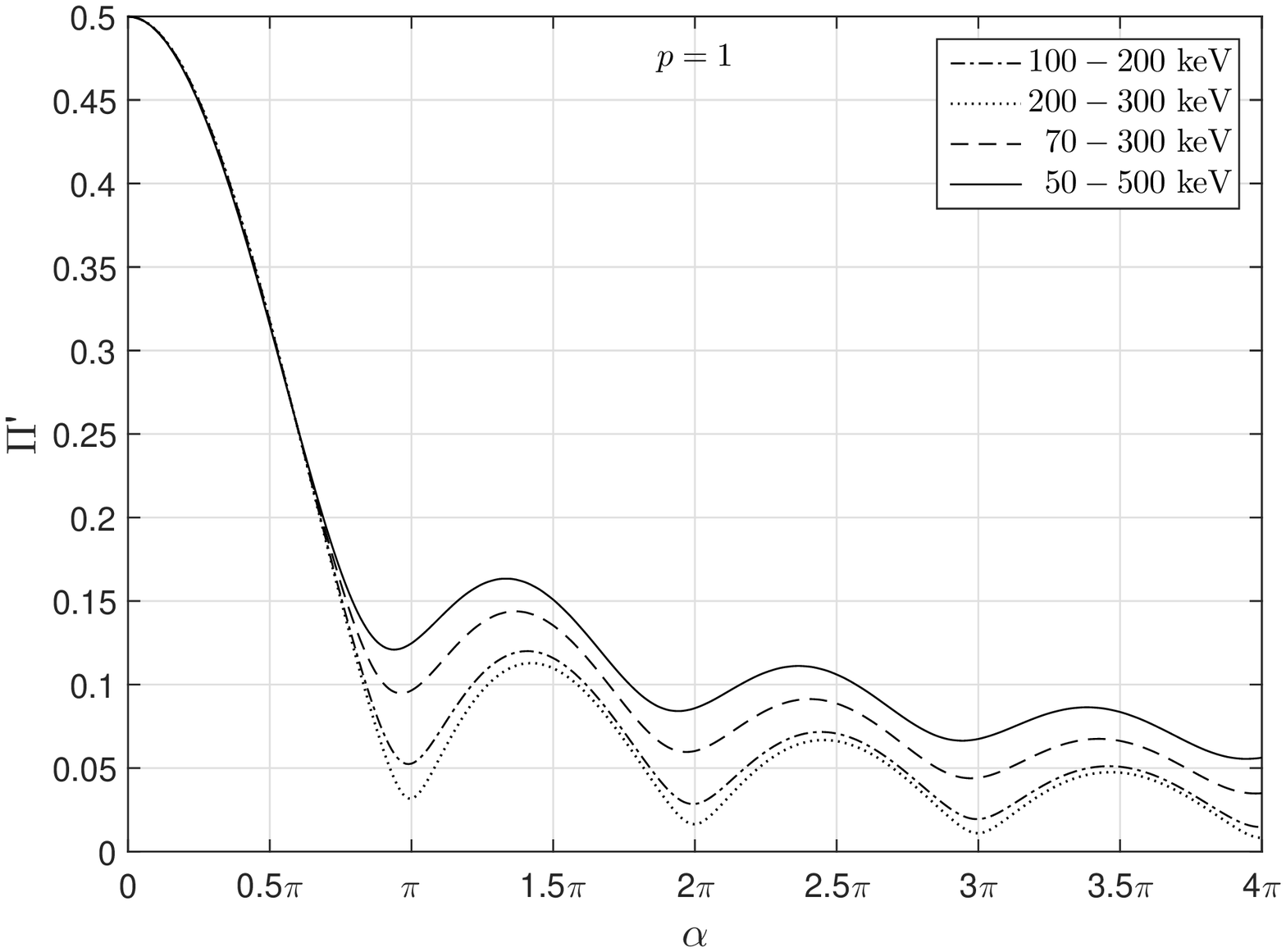}
 \includegraphics[width=0.5\textwidth]{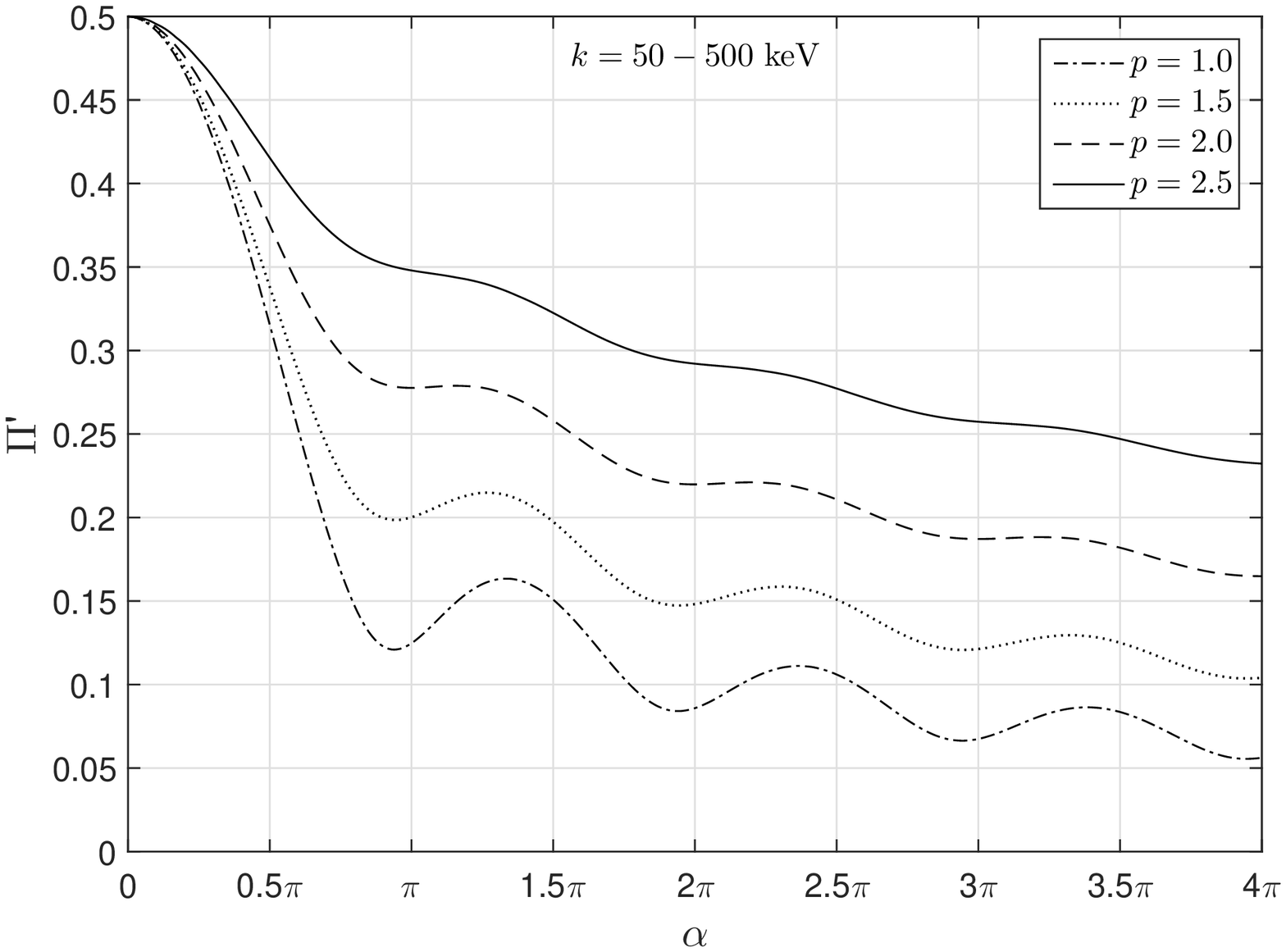}
 \caption{\small{Polarization degree as a function of $\alpha$ in the initially partially polarized case. Upper panel: the power-law index is fixed at $p=1$. Lower panel: the photons are in the energy band $k\in [50,500]$ keV.}}\label{fig:PI_alpha3}
\end{figure}

For any given $\alpha$, we numerically calculate $I'_{\rm max}$ and $I'_{\rm min}$, then calculate the polarization degree according to equation (\ref{eq:polarization2}). We plot the polarization degree as a function of $\alpha$ in Figure \ref{fig:PI_alpha3}.
In the upper panel, we plot the polarization for various energy bands, while the power-law index is fixed at $p=1$. In the lower panel, we show the polarization degree for various values of $p$, while the photons are assumed to be in the energy band $k\in[50,500]$ keV. Comparing with Figure \ref{fig:PI_alpha1}, we may see that the tendencies of polarization evolving with rotation angle in these two cases are very similar. Polarization degree as a function of $\alpha$ oscillates with a quasi-period $T\approx \pi$, and the oscillating amplitude decreases to zero gradually. Polarization has local minimums at $\alpha\approx n\pi$, and has local maximums at $\alpha\approx (n+1/2)\pi$. At $\alpha=\pi/2$, the polarization degree is larger than $30\%$. This is to say, more than $60\%$ of the initial polarization can be conserved at $\alpha=\pi/2$. The first local minimum locates at $\alpha\approx \pi$, whose value depends on the photon energy. At the same energy band, a hard spectrum has a larger net polarization. At the fixed rotation angle, photons in a wider energy band have larger polarization. All these futures are very similar to that in the initially completely polarized case. In fact, Figure \ref{fig:PI_alpha3} seems to be identical to Figure \ref{fig:PI_alpha1}, except that the ordinate is suppressed by a factor of $\sim 2$. This makes intuitive sense because the half polarized photons can be regarded as the mixture of $50\%$ completely polarized photons and $50\%$ completely unpolarized photons. The polarization degree of the initially completely polarized photons evolves according to Figure \ref{fig:PI_alpha1}, while the initially completely unpolarized photons keep unpolarized during the propagation.

\begin{figure}
 \centering
 \includegraphics[width=0.5\textwidth]{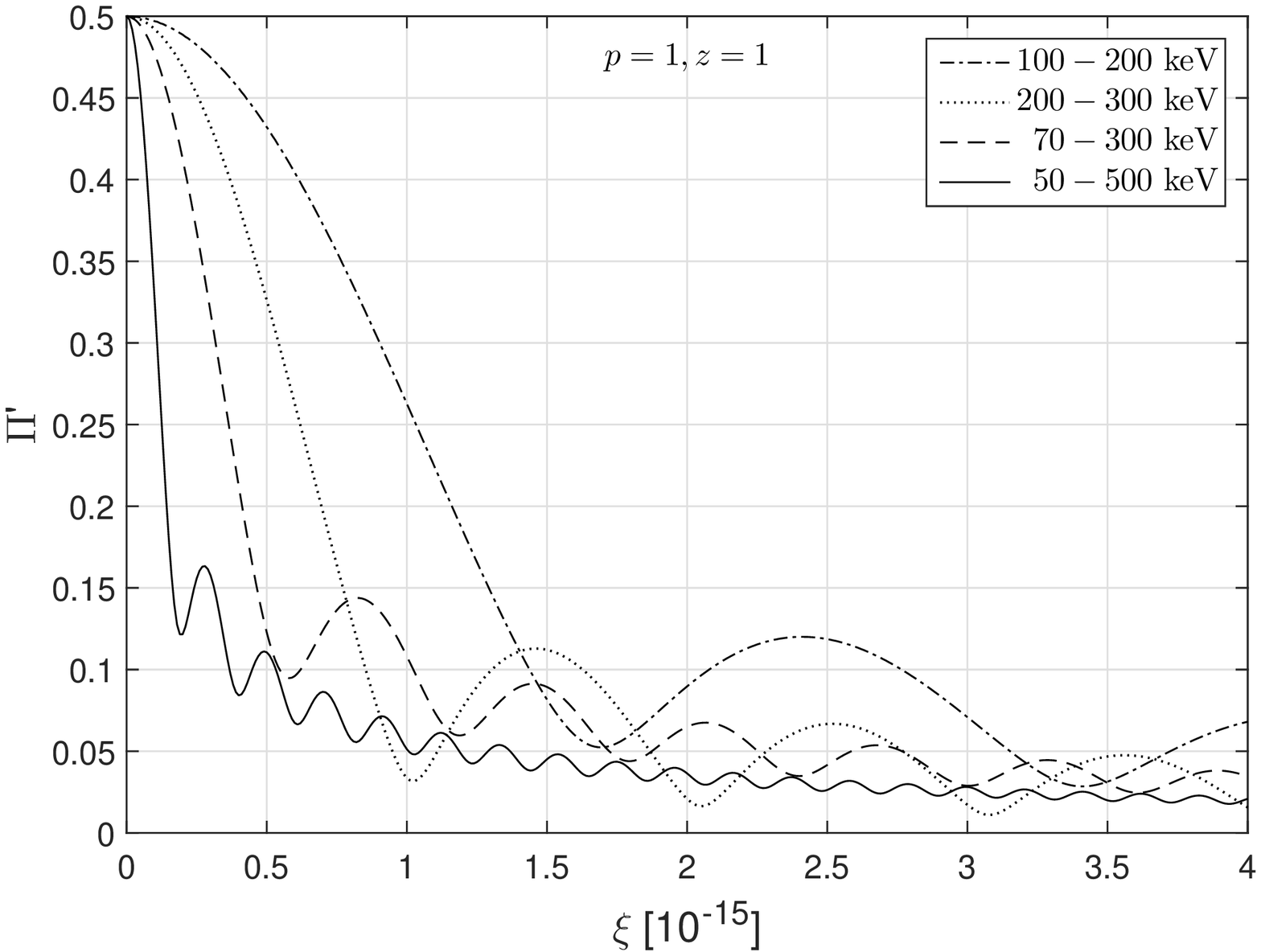}
 \includegraphics[width=0.5\textwidth]{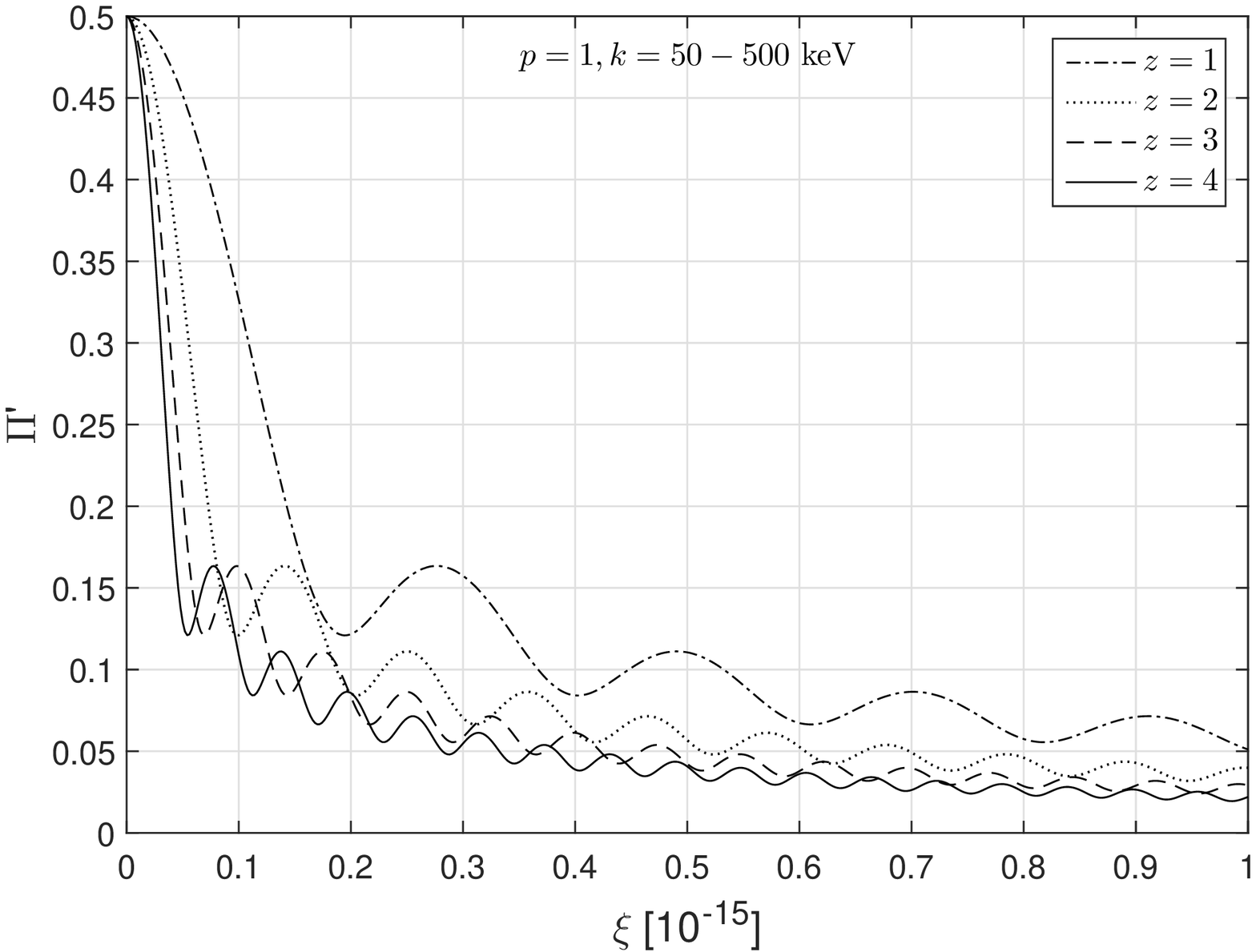}
 \caption{\small{Polarization degree as a function of $\xi$ in the initially partially polarized case. Upper panel: the redshift of source is $z=1$. Lower panel: the photons are in the energy band $k\in [50,500]$ keV. In both panels, the power-law index is $p=1$.}}\label{fig:PI_xi3}
\end{figure}

We also plot the polarization as a function of $\xi$ in Figure \ref{fig:PI_xi3}.
In the upper panel, the redshift of source is $z=1$, and different curves stand for different energy bands. In the lower panel, the photons are in the energy band $k\in [50,500]$ keV, and different curves stand for different redshifts of the source. In both panels, the power-law index is fixed at $p=1$. We may see that the evolution of polarization during the propagation is also similar to that in the initially completely polarized case. Firstly, the polarization degree decreases rapidly until reaching a local minimum as $\xi$ increases. The wider energy band has a steeper slope, but has a larger local minimum. After the first local minimum, the polarization degree oscillates with $\xi$, while the oscillating amplitude gradually vanishes.

\section{Discussions and Conclusions}\label{sec:conclusions}

In this paper, we have investigated the evolution of GRB polarization arising from LIV effect. The birefringence of light leads to the rotation of polarization vector duration propagation. We obtained the net polarization degree as a function of the rotation angle $\alpha$, where $\alpha$ represents the relative rotation angle of high-energy and low-energy photons. We showed that the net polarization degree decreases rapidly as $\alpha$ increases until $\alpha\approx \pi$. As $\alpha$ continuously increases, the polarization degree oscillates with a quasi-period $T\approx\pi$ and a gradually vanishing amplitude. More than $60\%$ of the intrinsic polarization degree can be conserved at $\alpha=\pi/2$. This is in conflict with the intuition that $\alpha$ couldn't be larger than $\pi/2$ when high polarization degree is observed. Hence, it is inappropriate to simply use $\pi/2$ as the upper limit to constrain LIV effect, especially when the photon energy band is wide and the spectrum is hard. Photons in a wider energy band have larger net polarization at the fixed $\alpha$. However, for a specific source, a wider energy band will also have a larger $\alpha$. The net effect is that photons in a wider energy band have a lower polarization degree. Therefore, the polarimetric observation in a wide energy band is favourable in constraining LIV. In addition, we found that GRBs with soft spectrum and high redshift are helpful to tightly constrain LIV. The compact space-borne Compton polarimeter {\it POLAR} onboard the Chinese space laboratory Tiangong-II is a high accuracy $\gamma$-ray polarimeter fully designed to measure the polarization of GRB in $50-500$ keV energy band. If a GRB at redshift $z\approx 1$ is observed by {\it POLAR} with $50\%$ polarization degree, and if the spectrum in the {\it POLAR} energy band follows the power-law distribution with index $p\approx 1$, then we can obtain the most conservative upper limit of LIV effect $\xi\lesssim 1\times 10^{-16}$. This is obtained by assuming that the GRB is intrinsically completely polarized. Otherwise, the constraint may be much tighter.

We apply our formulae to some true GRB events. \citet{Yonetoku:2012} claimed to have detected a polarization degree of $84_{-28}^{+16}\%$ in GRB 110721A in the {\it IKAROS-GAP} energy band $[70,300]$ keV. The photon spectrum in this energy band can be well fitted by the simple power law, with the power-law index $p=0.94\pm 0.02$ \citep{Tierney:2011}. Unfortunately, the redshift of this burst has not been directly measured. The $2\sigma$ lower limit of redshift inferred from the Amati relation is 0.45 \citep{Toma:2012}. Using these observational values, and assuming that photons are initially completely polarized, we obtain the upper limit of LIV effect $\xi\lesssim 4\times 10^{-16}$. GRB 061122 is a highly polarized GRB with redshift measurement $z=1.33$ \citep{Gotz:2013}. The polarization degree measured by the IBIS on board {\it INTEGRAL} in the energy band $[250,800]$ keV is $>60\%$ at $1\sigma$ confidence level \citep{Gotz:2013}. The spectrum can be fitted by a power law with an exponential cut-off, i.e. $N(k)\propto k^{-\alpha}\exp(-k/k_c)$, where $\alpha=1.15\pm 0.04$ and $k_c=221\pm 20$ keV. Using these observational parameters, the most conservative upper limit of LIV effect constrained from this burst is $\xi\lesssim 5\times 10^{-17}$.

Finally, it should be pointed out that in this paper, we only consider one type of LIV, i.e., doubly special relativity. The dispersion relation in equation (\ref{eq:dispersion}) breaks not only the Lorentz invariance, but also the CPT invariance. This case is not typically favored by theorists. It is widely discussed because it is one of the few theories of quantum gravity that can be tested. More importantly, the dispersion relation considered here is among the few which has the vacuum birefringence effect. If the vacuum birefringence effect does not exist, and the polarization vector does not rotate during the propagation, then the polarimetric observation couldn't be used to constrain LIV.

\section*{Acknowledgements}
We thanks the referee for the useful comments and suggestions to improve our paper. We are grateful to J. Li, H. Ma and L. Tang for useful discussions. This work has been supported by the Fundamental Research Funds for the Central Universities (Grant No. 106112016CDJCR301206), the National Natural Science Fund of China (Grant No. 11375203), and the Open Project Program of State Key Laboratory of Theoretical Physics, Institute of Theoretical Physics, Chinese Academy of Sciences, China (Grant No. Y5KF181CJ1).

\label{lastpage}

\end{document}